\documentclass[aps,pra,twocolumn,groupedaddress,floatfix]{revtex4-1}

\usepackage{amsmath}
\usepackage{bm}
\usepackage{graphics}
\usepackage{times}
\usepackage{graphicx}
\usepackage{xcolor}
\usepackage{array,multirow,graphicx}
\usepackage{multirow,hhline,graphicx,array}
\usepackage{rotating}
\usepackage{multirow}
\usepackage{booktabs}
\usepackage[markup=underlined]{changes}

\begin{document}

\title{Multi-sideband interference structures by high-order
   photon-induced continuum-continuum transitions in helium}
 
\author{D.~Bharti$^{1}$} 
\email{bharti@mpi-hd.mpg.de}
\author{ H.~Srinivas$^{1}$}
\author{F.~Shobeiry$^{1}$}
\author{A.~T.~Bondy$^{2,3}$}
\author{S.~Saha$^{3,4}$}
\author{K.~R.~Hamilton$^{5}$}
\author{R.~Moshammer$^{1}$}
\author{T.~Pfeifer$^{1}$}
\author{K.~Bartschat$^{3}$}
\author{A.~Harth$^{1,6}$} \email{Anne.Harth@hs-aalen.de}

\affiliation{$^1$Max-Planck-Institute for Nuclear Physics, D-69117 Heidelberg, Germany}
\affiliation{$^2$Department of Physics, University of Windsor, Windsor, ON N9B 3P4, Canada}
\affiliation{$^3$Department of Physics and Astronomy, Drake University, Des Moines, IA 50311, USA}
\affiliation{$^4$Department of Physics, Panskura Banamali College (Autonomous), Panskura, West Bengal 721152, India}
\affiliation{$^5$Department of Physics, University of Colorado at Denver, Denver, CO 80204, USA}
\affiliation{$^6$Center for Optical Technologies, Aalen University, D-73430 Aalen, Germany}

\date{\today}

\begin{abstract}
Following up on a previous paper on two-color photo\-ionization of Ar($3p$) [Bharti {\it et al.}, Phys. Rev. A~{\bf 103} (2021) 022834],
we present measurements and calculations for a modified three-sideband \hbox{(3-SB)} version of the ``reconstruction of attosecond beating by interference of two-photon transitions" \hbox{(RABBITT)} configuration applied to He($1s$).
The \hbox{3-SB} RABBITT approach allows us to explore interference effects between pathways involving different orders of transitions within the continuum. The relative differences in the retrieved oscillation phases of the three sidebands provide insights into the continuum-continuum transitions. The ground state of helium has zero orbital angular momentum, which simplifies the analysis of oscillation phases and their angle-dependence within the three sidebands.
We find qualitative agreement between our experimental results and the theoretical predictions for many cases but also observe some significant quantitative discrepancies.  
\end{abstract}

\maketitle

\section{Introduction}\label{sec:Intro}
The reconstruction of attosecond beating by interference of two-photon transitions \hbox{(RABBITT)} is a widely employed technique to characterize an attosecond pulse train and measure attosecond time delays in photo\-ionization processes, e.g.,~\cite{Paul2001,Muller2002,Klunder2011}. 
In the standard scheme, photoionization by various spectral harmonics in an attosecond extreme ultraviolet (XUV) pulse train (the pump photons) results in multiple discrete peaks (main peaks) within the photoelectron signal. Simultaneously, the presence of a time-delayed infrared (IR) field (the probe photons) creates an additional photoelectron peak between every two main peaks, referred to as ``sidebands." The photoelectron yield within these sidebands oscillates as a function of the time delay, and this oscillatory pattern can be utilized to determine the relative photoionization time delay.
However, in order to accurately determine photoionization time delays using the RABBITT method, it is essential to consider the contribution arising from the continuum-continuum transitions induced by the probe pulse.  This complexity is simplified by breaking down the phases of the two-photon transition matrix element into the sum of the Wigner phase, associated with the single-photon ionization process, and the continuum-continuum phase denoted as $\phi_{cc}$. Details can be found, for example, in Refs.~\cite{Dahlstr_m_2012, Dahlstr_m_2014,RevModPhys.87.765}. 
  
In 2019, Harth~{\it et al.}~\cite{Harth2019} introduced a variant of the RABBITT scheme known as three-sideband (3-SB) RABBITT, in which the interaction with the probe pulse results in the creation of not just one but three sidebands between every two main peaks.
In the context of the 3-SB RABBITT approach, Bharti~{\it et al.}~\cite{Bharti2021}  extended the ``decomposition approximation"  to determine the phase of the $N^{\rm th}$-order Above-Threshold-Ionization (ATI) matrix element. In this extension, the $\phi_{cc}$ contributions  from each participating transition are simply summed.
In the 3-SB scheme, one important consequence of the decomposition approximation is the prediction that the oscillation phases in the three sidebands formed between the same pair of main peaks should be identical, except for an additional~$\pi$ phase shift in the central sideband.
Numerical calculations performed for atomic hydrogen revealed slight deviations from this expectation, and these deviations consistently diminished with an increase in kinetic energy.  Experiments on atomic hydrogen, of course, are extremely challenging due to the difficulties in creating a dense H target.

A proof-of-principle experimental realization of the \hbox{3-SB} scheme was reported by Bharti~{\it et al.}~\cite{Bharti2023} for an argon target. As a noble gas with the first ionization potential of $\approx 15.8$~eV for the $3p$ electron, this target was experimentally favorable due to the fact that low-order (7, 9, 11, ...) 515~nm harmonics of the frequency-doubled fundamental field (1030~nm) in the XUV pulse train were able to ionize the atom. Additionally, ionization of the $3s$ electron is possible, resonances occur at relatively low ejected-electron energies of 10$-$15~eV due to the possible inner-shell promotion of the $3s$ electron, and the numerical treatment is very challenging. A further complication arises from the fact that already the XUV step alone promotes the $3p$ electron to two different angular-momentum states ($s$~and~$d$).  
All this, together with the uncertainties regarding the detailed time dependence of the electric fields, resulted in only qualitative agreement between the experimental data and theoretical predictions based on the \hbox{$R$-matrix} with time dependence (RMT) approach. 

In this paper, we report the outcomes of a 3-SB RABBITT experiment performed on \hbox{helium}.
The ground state of helium has zero orbital momentum, and hence the single-photon ionization induced by an XUV pulse results in the creation of a photo\-electron with a single orbital angular momentum~$\ell\!=\!1$. 
This simplifies the interpretation of the photo\-electron interference patterns resulting from the interaction of the photo\-electron with the IR photons. Due to its higher ionization potential of approximately 24.6~eV, however, helium presents further experimental challenges compared to argon, as it requires relatively high-energy XUV photons to create a sufficient number of main peaks. However, from a theoretical perspective, helium is significantly easier to handle than argon.  In fact, since the remaining $1s$ electron is tightly bound, a single-active-electron (SAE) approach may be sufficient to explain the basic features.

This paper is organized as follows. We begin with a brief review of the basic idea behind the \hbox{3-SB} setup in Sec.~\ref{sec:Scheme}. This is followed by a description of the experimental apparatus in Sec.~\ref{sec:Experiment} and the accompanying theoretical SAE and RMT approaches in Sec.~\ref{sec:Theory}.
We first show angle-integrated data
in Sec.~\ref{subsec:Angle-integrated}
before focusing on the angle-dependence of the \hbox{RABBITT} phases in the three sidebands of each individual group in 
Sec.~\ref{subsec:Angle-differential}.  
We finish with a summary and an outlook in Sec.~\ref{sec:Summary}.

\section{The 3-SB Scheme for Helium} \label{sec:Scheme}
 In this section, we briefly review the 3-SB scheme introduced in~\cite{Harth2019} and the analytical treatment presented in~\cite{Bharti2021}, as applied to the \hbox{3-SB} RABBITT experiment in general and then in our particular case of the helium target.

\begin{figure}[t!]
\includegraphics[width=1\columnwidth]{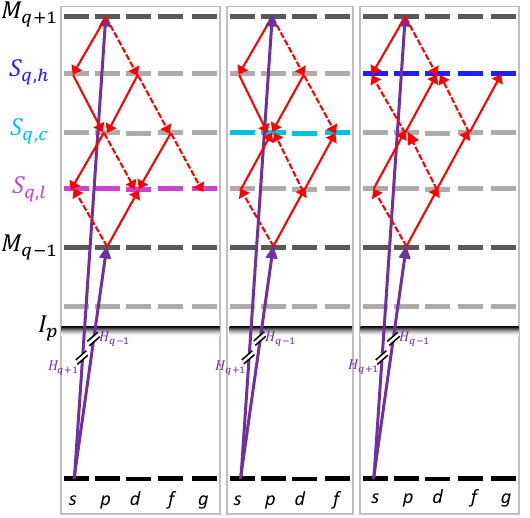}
\caption{3-SB transition diagram with all angular-momentum channels, illustrating only  the lowest-order pathways from emission and absorption processes necessary for yield oscillations in each sideband. 
$I_p$ is the ionization threshold, and the dashed line just above it labels the special case of the threshold sideband. See text for details. The solid and dashed lines in the transitions illustrate a propensity rule~\cite{PhysRevLett.123.133201,Bertolino_2020,Peschel2022}, with the solid line indicating a more probable transition compared to the dashed line. 
}
\label{fig:Fig5-10a}
\end{figure}

The basic scheme is illustrated in Fig.~\ref{fig:Fig5-10a}. 
The active electron originates from an $s$-orbital, and the absorption of an XUV photon with a frequency corresponding to either   $H_{q-1}$ or  $H_{q+1}$ transitions it to a $p$-orbital. This transition gives rise to the main photoelectron peaks labeled $M_{q-1}$ and $M_{q+1}$, respectively. The frequencies $H_{q-1}$ and $H_{q+1}$ correspond to the $(q\!-\!1)$ and $(q\!+\!1)$ odd harmonics, respectively, of the pulse used for generating the XUV train via high-order harmonic generation (HHG)~\cite{McPherson1987,Ferray1988}. Additional transitions within the continuum, resulting from either the absorption or emission of probe photons, lead to the emergence of three sidebands situated between two main peaks. These transitions to the sidebands can traverse different angular-momentum channels in compliance with the dipole selection rule.
We designate the trio of sidebands positioned between $M_{q-1}$ and $M_{q+1}$ according to their energy positions within the group, labeling them as $S_{q,l}$, $S_{q,c}$, and $S_{q,h}$, where the second subscript designates the lower, central, and higher sidebands, respectively.
All pathways leading to the same sideband interfere to produce the net photoelectron yield of the sideband. Changing the time delay between the XUV and IR pulses adds a dynamic phase to the interference of absorption and emission paths, leading to oscillations in the sideband yield.

As shown in~\cite{Bharti2023}, the general form of the signal in the sidebands is given by
\begin{align}
S_{q,j}(\tau,\theta) &=  A_{q,j}(\theta) + B_{q,j}(\theta) \, \cos(4\,\omega\tau -\phi_{R,q,j}(\theta)) 
\end{align} \label{eq:3Sbsiggnal}
with $j$ standing for $(l,c,h)$.
Each signal is characterized by a constant term~$A$, an oscillation amplitude~$B$, and a RABBITT phase~$\phi_{R}$.  As seen from the above equation, each of these parameters depends on the sideband group~$q$, the location of the sideband within that group~$(l,c,h)$, and the detection angle~$\theta$.  In angle-integrated detection mode, the angle-dependence is averaged over, but the general form of the equation remains unchanged. In both the angle-differential and angle-integrated cases, the signal oscillates as a function of the delay~$\tau$ with an angular frequency of four times the angular frequency~$\omega$ of the fundamental IR field.
Detailed expressions of the above parameters in terms of transition amplitudes can be found in~\cite{Bharti2021,Bharti2023}.  

\section{Experiment}\label{sec:Experiment}
Since the experimental methodology was described in detail in~Ref.~\cite{Bharti2023}, we only provide a brief summary of the most important settings for the current experiment. 
The fiber-based laser utilized in the experiment emits 50~fs (full width at half maximum) ultra-short infrared pulses centered around 1,030~nm with a pulse energy of 1.2~mJ.
 The pulse is divided into two arms of a Mach-Zehnder type interferometer using a holey mirror, which reflects approximately 85$\%$ of the beam in the pump arm and allows the remainder to pass through into the probe arm. Within the pump arm, a barium borate crystal is employed to produce the second harmonic (515~nm) of the laser pulse with an efficiency of 25-30$\%$.  
The fundamental IR beam is subsequently filtered via a dichroic mirror, and the second harmonic is focused onto an argon gas jet to produce an XUV pulse train through HHG. 
The driving 515~nm beam was spatially filtered out from the beamline, and the resulting XUV beam then passed through a 150~nm-thick aluminum filter.
Meanwhile, the IR beam within the probe arm was directed through a retro-reflector mounted on a piezo-electric-translation stage, before it was spatially and temporally recombined with the XUV beam. Both the XUV and IR beams were focused inside a reaction microscope (ReMi)~\cite{Ullrich_2003} onto a supersonic gas jet of the target species.

The relative strength of the harmonics was determined by analyzing the photoelectron spectra generated solely by the XUV beam in different gases. Beyond the helium ionization threshold, four harmonics ($H_{11}$, $H_{13}$, $H_{15}$, and $H_{17}$) were observed, with their strength decreasing as the photon energy increased.
Below the ionization threshold, $H_7$ is positioned at the low-energy transmission edge of an aluminum filter, resulting in significant attenuation of this harmonic and effectively elimination of all harmonics below $H_7$. The harmonic $H_{9}$, whose energy is also below the ionization energy, is crucial for generating oscillations in the yield of the threshold sideband (see the dashed line just above the ionization threshold in Fig.~\ref{fig:Fig5-10a}). This threshold sideband ($S_{th}$) belongs to the so-called ``uRABBITT'' \hbox{scheme~\cite{Swoboda2010, Galan2014, Villeneuve2017, atoms9030066, Autuori2022, Drescher2022,Neoricic2022}}.

We adjusted the IR probe intensity by utilizing an iris in the probe arm. 
\hbox{RABBITT} measurements were performed at three IR probe peak intensities: $I_1 \approx 5 \times \mathrm{10^{11}~W/cm^2}$ , \hbox{$I_2 \approx 7 \times \mathrm{10^{11}~W/cm^2}$}, and $I_3 \approx 1.2 \times \mathrm{10^{12}~W/cm^2}$. 
The IR and XUV fields were linearly polarized parallel to each other and along the spectrometer axis of the ReMi. 
The latter enables the reconstruction of the three-dimensional momenta of the electrons and corresponding ions (in coincidence) created during the photoionization process~\cite{Ullrich_2003}. Furthermore, the ReMi is capable to select electrons emitted in different directions, thereby enabling the construction of photo\-electron angular distributions (PADs).

The XUV-IR  temporal delay was sampled at regular intervals of $T_0/60$, where $T_0=3.44$~fs is the period of the IR pulse. The 
delay was scanned over a range equivalent to one and a half times the optical cycle of the IR pulse. 
The XUV-IR beamline was actively stabilized~\cite{Srinivas:22, Bharti2023} to achieve a stability of approximately 30 attoseconds over a data acquisition period of more than ten hours.

\section{Theory}\label{sec:Theory}
The numerical approaches to model the experiment were also described in detail in previous publications.  Hence, we again limit ourselves to a brief summary with references given at the appropriate spots.

\subsection{The SAE approach}\label{subsec:SAE}
We employed the same SAE model as Birk {\it et al.}~\cite{Birk2020} and Meister {\it et al.}~\cite{PhysRevA.102.062809}. 
Specifically, we used the one-electron potential 
\begin{equation}
V(r) = -\frac{1}{r} - \left(\frac{1}{r} + 1.3313\right) \,\exp(-3.0634\,r),
\end{equation}
where $r$ is the distance from the nucleus, to calculate the valence orbitals. The difference of excitation
energies compared to the recommended excitation from the NIST database~\cite{NIST} is less than
0.2$\,$eV even in the worst-case scenario.

For both the XUV pulse train and the fundamental IR, we used temporal fields based on the measured spectrum of the IR pulse and on the measured XUV-only photoionization spectrum considering the experimental resolution.  This gives reasonably accurate information about the relative strength of the harmonics. Additionally, we incorporated an estimated atto-chirp into the theoretical XUV pulse train.

Since both the XUV pulse train and the fundamental IR are linearly polarized along the same direction, the initial state can be propagated very efficiently and accurately. Specifically, we used an updated version of the code described by Douguet
{\it et al.}~\cite{PhysRevA.93.033402}. In contrast to the heavier noble gases, the SAE approach is expected to be suitable for the helium target, as long as obvious two-electron correlation effects, e.g., auto\-ionizing resonances, are not affecting the process significantly. This is, indeed, the case for the present study. 

\subsection{The RMT approach}\label{subsec:RMT}
As a second method, we employed the general \hbox{$R$-matrix} with time dependence (RMT)
method~\cite{BROWN2020107062}.
To calculate the necessary time-independent basis functions and dipole matrix elements for the present work, we set up the simplest possible model, a non\-relativistic \hbox{1-state} 
approach. This model, labeled \hbox{RMT} below, enables efficient calculations whose predictions can be readily compared with both the experimental data and those from the SAE approach to make a first assessment regarding the likely quality of the theoretical predictions.

We took the same pulse as in the SAE calculation.  Instead of the numerical orbital employed in the SAE calculation, which is close to the Hartree-Fock orbital of the ground-state configuration, however, we took the known $1s$ orbital of He$^+$.  While this is not optimal to obtain the very best ground-state energy, this disadvantage is mitigated almost completely by the continuum-continuum terms in the \hbox{$R$-matrix} hamiltonian. Using this orbital will, however, be advantageous in future studies, where we plan to include additional states in the close-coupling expansion to check the convergence and sensitivity of the numerical predictions. In contrast to SAE, exchange effects between the two electrons are treated explicitly within the \hbox{$R$-matrix} box.  

\section{Results and Discussion}\label{sec:Results}
We first present in Sec.~\ref{subsec:Angle-integrated} our angle-integrated \hbox{RABBITT} results. This is followed by a discussion of angle-differential measurements in Sec.~\ref{subsec:Angle-differential}. 
For the latter case, we will concentrate on only two sideband groups, $S_{12}$ and $S_{14}$, where we have a sufficient amount of data  to conduct  the angle-resolved investigation of the RABBITT phase. 

\subsection{Angle-Integrated RABBITT scans}\label{subsec:Angle-integrated}

\begin{figure}[t!]
\includegraphics[width=0.98\columnwidth]{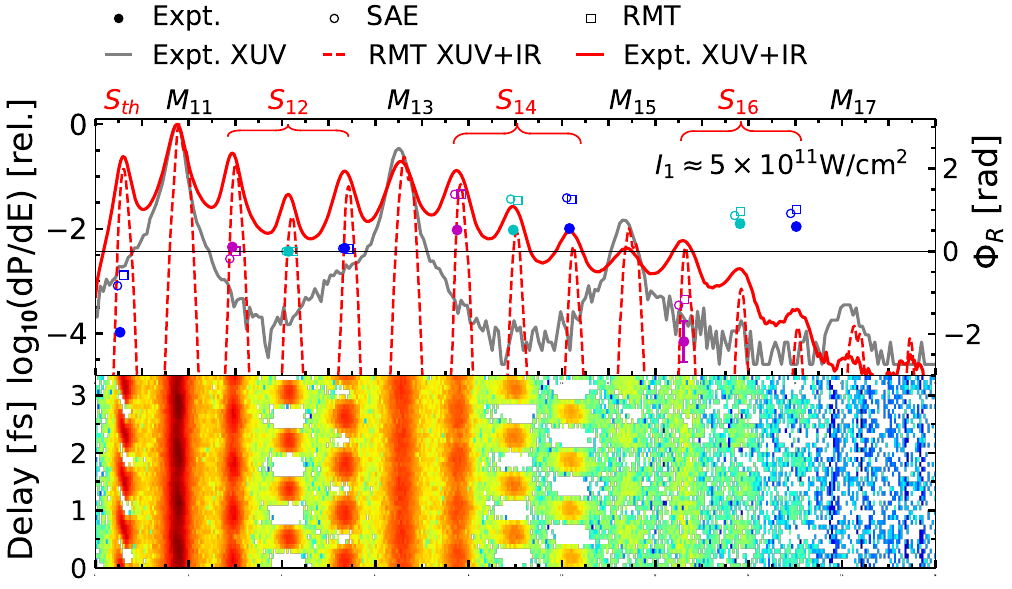}
\includegraphics[width=0.98\columnwidth]{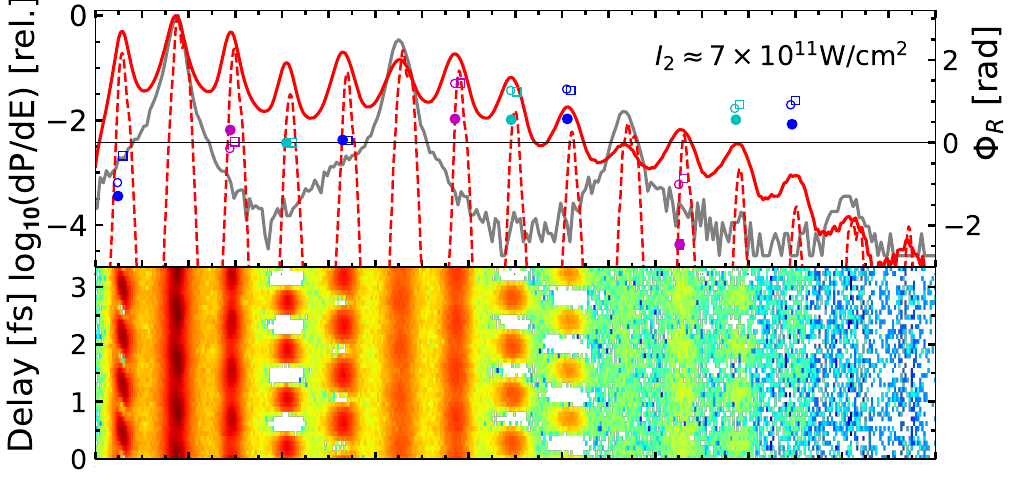}
\includegraphics[width=0.98\columnwidth]{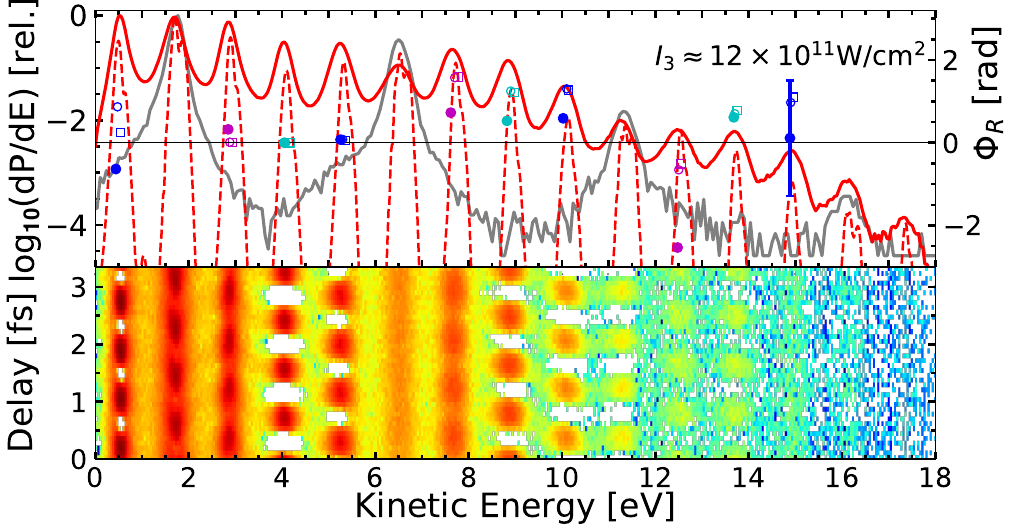}
\caption{Results from the angle-integrated \hbox{RABBITT} measurements taken at peak IR intensities $I_1 \approx \rm 5 \times 10^{11}W/cm^2$~(top), $I_2 \approx \rm 7 \times 10^{11}W/cm^2$~(center), and $I_3 \approx \rm 12 \times 12^{11}W/cm^2$~(bottom). The logarithmic colormap in each of the lower panels displays the angle-integrated \hbox{RABBITT} traces, while the upper panel shows the XUV-only (gray line) and the delay-integrated photo\-electron spectrum (red line) normalized to the peak value. The dashed line in the top panels is the (relative) signal predicted by the RMT model. The main lines and the sideband groups are also indicated.
The right $y$-axis shows the phases of the delay-dependent yield oscillations of the sidebands obtained from our fitting method, along with their estimated fitting errors. Since the absolute phase is not known experimentally, we set it to zero for the center sideband in the $S_{12}$ group. Also, a phase~$\pm\pi$ was added to the center sideband to simplify the comparison with the other two.  The solid circles represent the experimental data while the open symbols are the theoretical SAE (circles) and RMT (boxes) predictions.
}
\label{fig:Fig5-3}
\end{figure}

Figure \ref{fig:Fig5-3} shows the results from three angle-integrated \hbox{RABBITT} measurements recorded at the probe intensities $I_1 \approx \rm 5 \times 10^{11}W/cm^2$~(top), $I_2 \approx \rm 7 \times 10^{11}W/cm^2$~(center), and $I_3 \approx \rm 12 \times 12^{11}W/cm^2$~(bottom). 
The log-scale colormap in the lower part of each panel shows the \hbox{RABBITT} trace obtained by subtracting the delay-integrated signal from the original trace to highlight the oscillations.
The upper part of the panels displays the XUV-only photo\-electron spectrum (gray line) and the delay-integrated photo\-electron spectrum (red line), both normalized to their peak values and plotted on a logarithmic scale.

Looking at the \hbox{RABBITT} traces, we see that the central sideband displays the clearest oscillation at all applied IR  intensities. This was expected since both the absorption and the emission paths that populate this sideband are of the same order.  In the lower and higher sidebands, the two interfering terms are of different (second and fourth) orders, and hence the contrast in the oscillation is reduced  compared to the central sideband. 

Next, the \hbox{RABBITT} traces demonstrate that the oscillation contrast is generally better for the higher sideband in each group than for the lower one. This is due to the rapid decrease of the main peaks with increasing energy, which makes the amplitude of the two transition paths in the interference more balanced in the case of the higher sideband.
In $S_{12,h}$, for example, the upper main peak ($M_{13}$) is much weaker than the lower main peak ($M_{11}$), resulting in the magnitude of a three-photon transition from the stronger lower main peak \hbox{($M_{11}+3\,\omega$)} becoming comparable to the magnitude of a one-photon transition from the weaker upper main peak ($M_{13}-\omega$). Hence the interference of these two terms leads to a strong delay-dependent oscillation. In contrast, for the lower sideband $S_{12,l}$, the magnitude of a one-photon transition from the lower main peak \hbox{($M_{11}+\omega$)} is much stronger than the magnitude of a three-photon transition from the already weaker upper main peak \hbox{($M_{13}-3\,\omega$)}, thus resulting in small contrast of the oscillation.

Additionally, at the lowest applied IR intensity (top panel of Fig.~\ref{fig:Fig5-3}), the highest energy main peak $M_{17}$ is almost entirely depleted, but it becomes repopulated as the intensity is increased. The same is seen for the main peak $M_{15}$, which first becomes weaker in the presence of the IR pulse but then gains strength when the intensity is increased.
Looking at the $S_{16}$ group, we observe that the contrast in the lower sideband ($S_{16,l}$) is very weak at the lowest applied intensity, but it gradually improves as the IR intensity increases. On the other hand, the contrast of the oscillation in the higher sideband $S_{16,h}$ deteriorates with increasing IR intensity. 

\begin{figure}[!t]
\includegraphics[width=0.95\columnwidth]{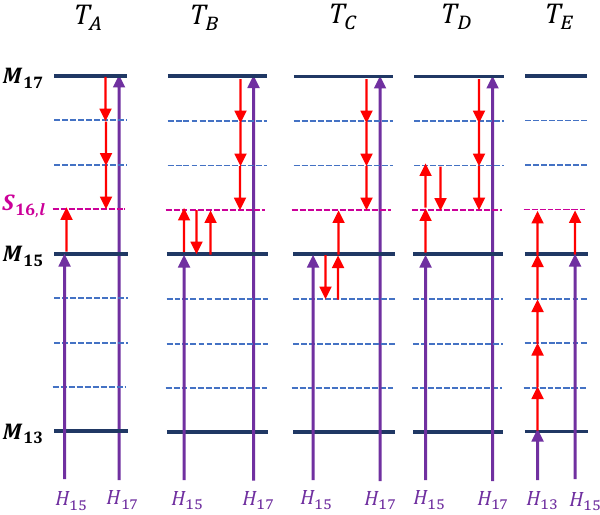}
\caption{Selected interference schemes for $4 \omega$ oscillations in the lower sideband. Scheme $T_A$ dominates at low probe intensity. The oscillation generated by scheme $T_A$ is out of phase by $\pi$ compared to the oscillations produced by schemes $T_B$, $T_C$, $T_D$, and $T_E$.}
\label{fig:Fig5-4}
\end{figure}

\begin{table*}
    \begin{tabular}{|c|c|c|c|c|c|c|c|c|c|c|}
    \hline
        probe & $S_{10}~(S_{th}$) & \multicolumn{3}{|c|}{$S_{12}$} &  \multicolumn{3}{|c|}{$S_{14}$}&  \multicolumn{3}{|c|}{$S_{16}$} \\ 
        \cline{2-11}
        ~intensity~ & $S_h$ & $S_l$ & $S_c$ & $S_h$ & $S_l$ & $S_c$ & $S_h$ & $S_l$ & $S_c$ & $S_h$ \\ \hline
        \multirow{2}{*}{$I_1$} &$-1.96$& $0.11$ & $0.00$ ~&~ $0.08$ & $0.52$ & $0.52$ & $0.55$ & $-2.18$ & $0.67$ & $0.60$ \\ 
        &~~ $\pm 0.02$ ~~&~~ $\pm 0.02$ ~~&~~ $\pm 0.01$ ~~&~~ $\pm 0.02$ ~~&~~ $\pm 0.06$ ~~&~~ $\pm 0.01$ ~~&~~ $\pm 0.01$ ~~&~~ $\pm 0.49$ ~~&~~ $0.08$ ~~&~~ $\pm 0.06$ ~~\\ \hline
        \multirow{2}{*}{$I_2$}  & $-1.29$ & $0.31$ & $0.00$ & $0.07$ & $0.58$ & $0.56$ & $0.58$ & $-2.46$ & $0.56$ & $0.54$ \\ 
        & $\pm 0.01$ & $\pm 0.02$ & $\pm 0.01$ & $\pm 0.01$ & $\pm 0.04$ & $\pm 0.01$ & $\pm 0.01$ & $\pm 0.11$ & $0.04$ & $\pm 0.07$ \\ \hline 
        \multirow{2}{*}{$I_3$}  & $-0.64$ & $0.32$ & $0.00$ & $0.07$ & $0.73$ & $0.53$ & $0.59$ & $-2.53$ & $0.61$ & $0.11$ \\ 
        & $\pm 0.01$ & $\pm 0.02$ & $\pm 0.01$ & $\pm 0.01$ & $\pm 0.02$ & $\pm 0.11$ & $\pm 0.01$ & $\pm 0.05$ & $0.03$ & $\pm 1.40$ \\ \hline 
    \end{tabular}
\caption{Angle-integrated \hbox{RABBITT} phase (in radians) obtained from the fitting procedure to the measurements at three IR intensities.
}
\label{tab:Tab5-1}
\end{table*}

When the kinetic energy or the IR intensity is increased, higher-order transitions should be accounted for as well.  
Figure~\ref{fig:Fig5-4} shows a selection of many-order transition pathways leading to the lower sideband of the $S_{16}$ group and the interference schemes contributing to the oscillations of the yield. 
At the low  IR intensity, the oscillation in the photo\-electron yield of the lower  sideband  is predominantly influenced by the interference scheme $T_A$.
However, 
with increasing kinetic energy or IR intensity, the involvement of higher-order transitions featuring $M_{15}$ and $M_{13}$  becomes significant in shaping the oscillations of the yield.

In order to determine the phases of the oscillation from each sideband, the photo\-electron spectrum was first integrated within an energy window of 0.8~eV across the peak and was then fitted to a cosine function of the form \hbox{$A+B\,\mathrm{cos}\,(4\,\omega \tau -\phi_R)$}. For the lower and higher sidebands, the trivial $\pi$ phase was removed from the obtained phases. As the absolute experimental phase is unknown, the retrieved phases from the three measurements were shifted to align the data point for $S_{12,c}$ at zero for comparison of the relative phases. The  phases of the oscillations obtained from the fitting process, along with their corresponding fitting errors, are shown in the upper panels of Fig.~\ref{fig:Fig5-3}  and also listed in Table~\ref{tab:Tab5-1}.

\begin{figure*}
\includegraphics[width=0.6\columnwidth]{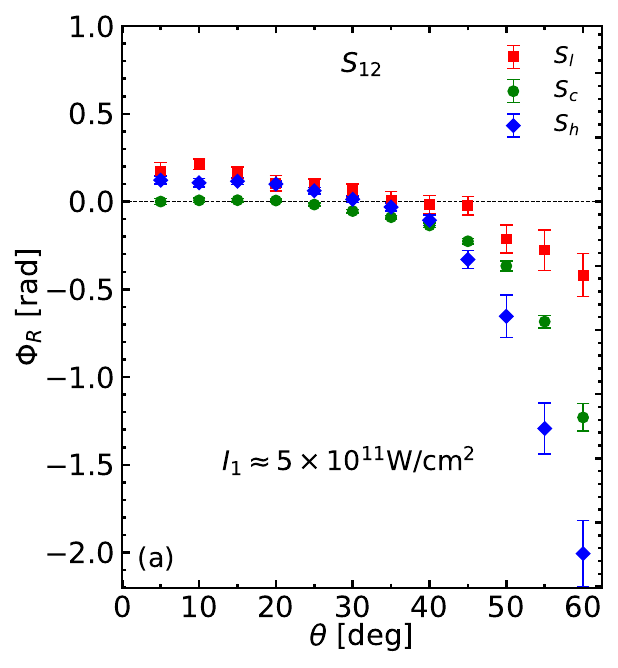}
\includegraphics[width=0.504\columnwidth]{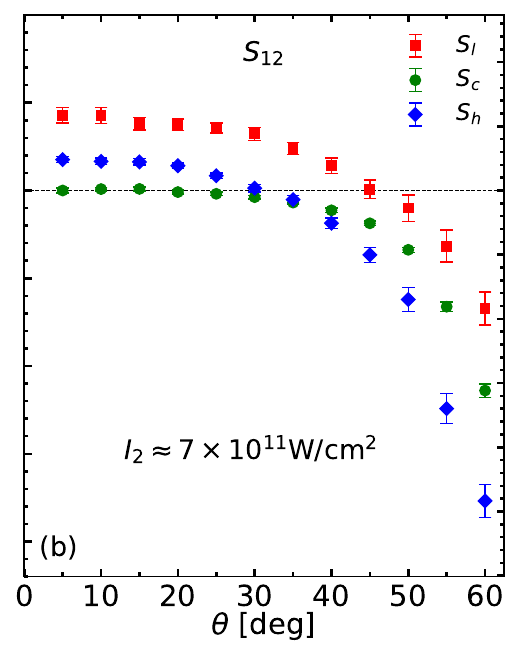}
\includegraphics[width=0.618\columnwidth]{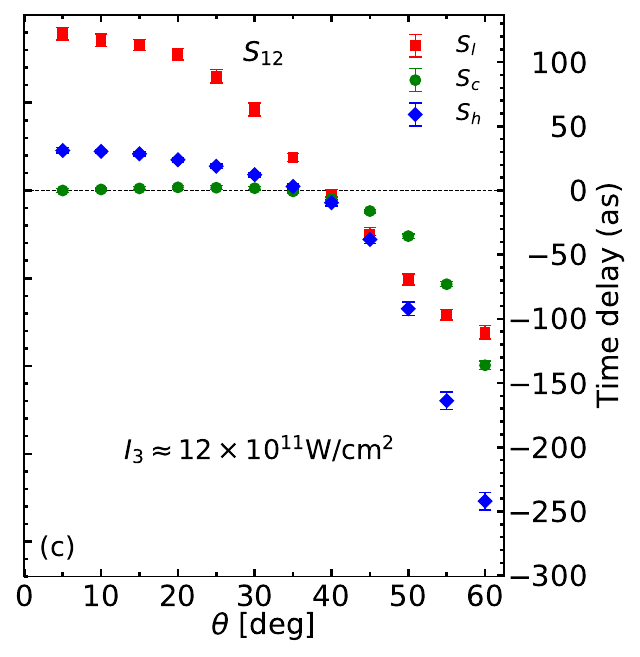}
\caption{Angle-dependent phase retrieved from the $S_{12}$ group in the \hbox{RABBITT} scans for IR peak intensities of $I_1 \approx \rm 5 \times 10^{11}W/cm^2$~(a), $I_2 \approx \rm 7 \times 10^{11}W/cm^2$~(b), and $I_3 \approx \rm 12 \times 10^{11}W/cm^2$~(c).}
\label{fig:Fig5-8}
\end{figure*}

\begin{figure*}
\includegraphics[width=0.6\columnwidth]{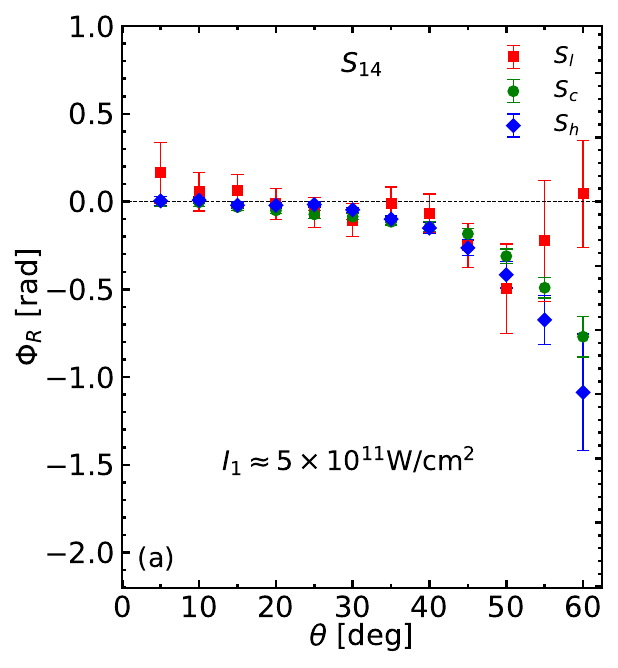}
\includegraphics[width=0.504\columnwidth]{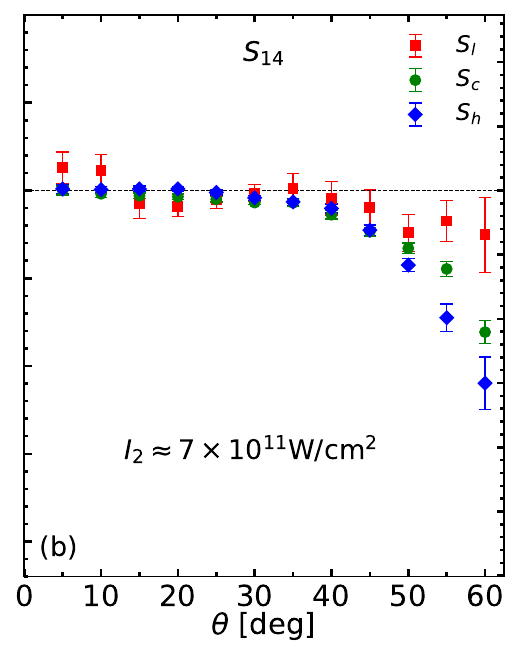}
\includegraphics[width=0.618\columnwidth]{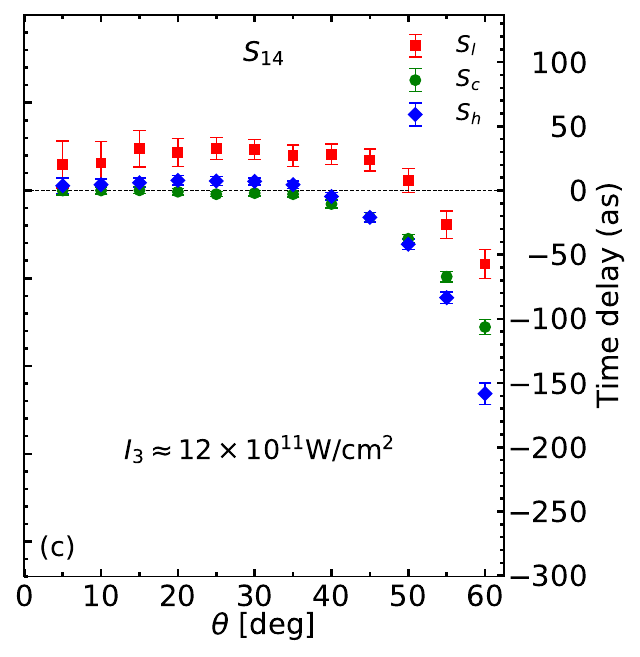}
\caption{Angle-dependent phase retrieved from the $S_{14}$ group in the \hbox{RABBITT} scans for IR peak intensities of $I_1$ (a), $I_2$ (b), and $I_3$ (c).}
\label{fig:Fig5-9}
\end{figure*}

Several noteworthy results can be seen from Table~\ref{tab:Tab5-1}: 1)~After accounting for the additional phase of~$\pi$, the phases in the $S_{12}$ and $S_{14}$ groups are nearly the same in all three sidebands of the respective group.  This is in excellent agreement with the ``decomposition approximation''~\cite{Bharti2021}. 2)~Relative to $S_{12}$, the phases in the $S_{14}$ group are larger by~$\approx 0.5$~rad. 3)~The lower sideband in $S_{16}$ exhibits a significantly different phase compared to the other two. This is likely due to the contribution of the higher-order interference schemes ($T_B, T_C,$  $T_D,$ and $T_E$ in Fig.~\ref{fig:Fig5-4}) dominating over the lowest-order interference ($T_A$). 
Since these interference schemes, involving higher-order pathways, produce oscillations at the same frequency but $\pi$ out of phase with that arising from the lowest-order interference scheme $T_A$,  their contribution in the yield oscillation results in the observed change \hbox{of~$\approx \pi$}. It should also be noted that the oscillation due to $T_E$ includes the spectral-phase difference of $H_{15}$ and $H_{13}$, while the schemes $T_A,$ $T_B,$ $T_C,$ and $T_D$ contain the spectral phase difference of $H_{17}$ and $H_{15}$. 

Regarding the agreement between experiment and theory, we see that both SAE and RMT agree with the experimental findings of nearly
the same phases for each member of the $S_{12}$ and $S_{14}$ groups.
Furthermore, the theoretical phases within $S_{14}$ are systematically larger than those observed in the experiment. This is simply due to an overestimate of the chirp that was included in the description of the XUV pulse. Since RABBITT calculations are very time-consuming, we decided not to repeat them with only the chirp being reduced.  There is also good qualitative agreement in the fact that $S_{16,c}$ and $S_{16,h}$ have about the same RABBITT phase, while that of $S_{16,l}$ is very different. 

Finally, there is significant quantitative disagreement between experiment and the two theories, as well as between the two theoretical predictions themselves, for the threshold sideband~$S_{th}$, which is the upper member of the $S_{10}$ group and the only one above the ionization threshold.  However, there is general agreement in the qualitative finding of a strong intensity dependence.  This strong sensitivity to the intensity is caused by near-resonant interactions with Rydberg states. 

\subsection{Angle-Differential RABBITT scans}\label{subsec:Angle-differential}
To examine the angle-dependence of the \hbox{RABBITT} phases, we analyzed the photo\-electrons emitted at various angles relative to the polarization axis of the driving light fields. This involved segregating the data into angle-differential data sets, which were generated by integrating the photo\-electron yield over 10-degree angular windows at 5-degree intervals.
The phase of the oscillation in each sideband was extracted from each angle-differential data set using the same method as outlined in the angle-integrated case. The photo\-electron spectra from each sideband were integrated over a 0.6 eV energy window centered on the peak, and the resulting delay-dependent signal was fitted to a cosine function to obtain the phase.

Figures \ref{fig:Fig5-8} and \ref{fig:Fig5-9} illustrate the angle-dependence of the retrieved phases from the three sidebands in the $S_{12}$ and $S_{14}$ groups, respectively, for different IR intensities: (a) $I_1$, (b)~$I_2$, and (c)~$I_3$. For better comparison and since we do not know the absolute experimental phase, a common phase was added to the three sidebands of each group to shift the central sideband phase to zero at the first data point ($5^\circ$) for each intensity.
The statistical uncertainty of the data above $60^\circ$ was insufficient to retrieve a meaningful oscillation phase at all, and the estimated fitting errors suggest some caution regarding the extracted phases at the larger angles, especially at the lowest intensity. 
The figures demonstrate a consistent angle-dependent RABBITT phase in the central sidebands of both the $S_{12}$ and $S_{14}$ groups across all three probe intensities. In contrast, the lower and the higher sidebands exhibit some dependence on both their group and the IR intensity.

We first discuss the angle-dependent oscillation phase in the central sideband while considering only the lowest-order transitions depicted in Fig.~\ref{fig:Fig5-10a}. 
According to the selection rules for electric dipole transitions, a three-photon transition to the central sideband results in the creation of a combination of two partial waves corresponding to angular momentum states of $p$~$(\ell \!=\! 1$) and $f$~$(\ell \!=\! 3$), where $m=0$ is conserved.  
The amplitudes of these partial waves vary with the angle of electron emission and are determined by the spherical harmonics $Y_{\ell,0}(\theta)$.   In our specific experimental setup, where both beams are linearly polarized along the same direction, we can substitute the spherical harmonics with Legendre polynomials denoted as $P_{\ell}(\theta)$.
Consequently, the interference of all partial waves generated during the absorption and emission processes will result in a signal of the form
\begin{subequations}
\begin{align}
S_c(\tau,\theta) &=  A_{c}(\theta) + B_{c}(\theta) \, \cos(4\,\omega\tau -\phi_{R,c}(\theta))\\
&=A_{c}(\theta) + a_{11} P_1^2(\theta)\,\cos(4\,\omega\tau -\phi_{11})\\
&+a_{33} P_3^2(\theta)\,\cos(4\,\omega\tau -\phi_{33}) \\
   & + P_1(\theta)\,P_3(\theta)\,\big[ a_{13}\,\cos(4\,\omega\tau -\phi_{13}) \nonumber\\
    &\qquad ~~~~~~~~~~~~~~~+a_{31}\,\cos(4\,\omega\tau -\phi_{31}) \big].
\end{align} \label{eq:test}
\end{subequations}
Here, $A_{c}(\theta)$ represents the signal that does not oscillate during the temporal delay scan. Each oscillation term is characterized by an amplitude determined by the magnitude of the dipole transition matrix elements~$a_{\ell,\ell'}$ and products of the Legendre polynomials $P_{\ell}(\theta)P_{\ell'}(\theta)$, as well as a phase~$\phi_{\ell,\ell'}$. Here, $\ell$ and $\ell'$ span across the possible orbital angular-momentum quantum numbers reached in the absorption and emission processes, respectively.
Referring to Eqs.~(3), we use the term ``same-channel interference" (3b,3c) to indicate interference occurring between two partial waves with the same angular momentum ($\ell \!= \!\ell'$) and ``cross-channel interference" (3d,3e) to indicate interference between two partial waves with different angular momenta ($\ell \!\neq \!\ell'$).

The oscillation phase~$\phi_{\ell,\ell'}$ of each interference term in Eq.~(\ref{eq:test}) includes the XUV chirp, the Wigner phases, and phases arising from continuum-continuum transitions. While the XUV chirp and Wigner phases remain constant across all $\phi_{\ell,\ell'}$, the contribution of the continuum-continuum coupling phase ($\phi_{cc}$) varies based on the angular momentum of the states involved in the transition pathways.
However, the variability in $\phi_{cc}$ across distinct angular momenta is typically small, usually less than $\pi/10$ for kinetic energies exceeding 5~eV. Moreover, this variability diminishes  with increasing kinetic energy~\cite{Peschel2022}. Consequently, differences among the oscillation phases~$\phi_{\ell,\ell'}$ across various interference terms are generally small, decreasing further with increasing kinetic energies.

Due to the distinct oscillation phase $\phi_{\ell,\ell'}$ in each of the interfering terms and the variation in the associated oscillation amplitude, which contains the Legendre polynomials, with the angle of electron emission, the overall retrieved oscillation phase $\phi_{R}(\theta)$ is angle-dependent~\cite{Heuser2016, Fuchs:s}.
A significant change in $\phi_{R}(\theta)$ may occur when the electron emission angle gets close to the angle where the cross-channel interference amplitude equals the combined amplitude of other interference terms. 
In the cross-channel interference expressions within Eqs.~(3), the product $P_{\ell}(\theta)P_{\ell'}(\theta)$ changes sign upon passing through a node in one of the Legendre polynomials. This sign change is effectively equivalent to adding a $\pi$ phase in the oscillation phase $\phi_{\ell,\ell'}$.  This phase addition results in a sudden change in the overall RABBITT phase $\phi_{R}$ around the angle where the cross-channel's amplitude surpasses that of the same-channel interference terms.
The drop in the \hbox{RABBITT} phase $\phi_{R}(\theta)$ in the central sideband in Fig.~\ref{fig:Fig5-8}(a) above $40^\circ$ suggests that the angle-dependent amplitude of the cross-channel ($p\!-\!f$) interference is larger than that of the same-channel ($p\!-\!p$ and $f\!-\!f$) interferences.
Since the angle-dependent behavior of the RABBITT phase for the central sideband remains consistent across all three applied probe intensities, the aforementioned arguments apply to all three cases.

In the subsequent sideband group $S_{14}$, the discrepancy in the oscillation phase $\phi_{\ell,\ell'}$ among various interfering terms reduces due to the further decreased variation in $\phi_{cc}$ between different angular momenta. This leads to an almost negligible angular dependence of the RABBITT phase until the emission angle approaches a critical point where the dominance of cross-channel interference over other interfering terms triggers an abrupt $\pi$ phase shift. 
While this is not seen in the angular regime for which we have experimental data, it is very noticeable in the theoretical predictions for the larger angles (cf.\ Fig.~\ref{fig:Fig-xx} below).

We next discuss the angular variation of the RABBITT phase in the lower and higher sidebands of $S_{12}$ for the case of the lowest applied intensity, as shown in Fig.~\ref{fig:Fig5-8}(a). 
In both the lower and higher sidebands, a four-photon transition populates $s$, $d$, and $g$ states, while a two-photon process populates $s$ and $d$ states (cf.\ Fig.~\ref{fig:Fig5-10a}). 

Based on the propensity rule for continuum-continuum transitions~\cite{PhysRevLett.123.133201,Bertolino_2020,Peschel2022}, the creation of a $g$ electron  in the lower sideband via emission of three IR photons is less likely. Consequently, despite the occurrence of a $\pi$-jump in the oscillation phases of $s\!-\!g$ and $d\!-\!g$ cross-channel interferences near $30^\circ$, this does not result in a significant change in the RABBITT phase due to the reduced oscillation amplitude associated with cross-channel interferences involving a $g$~electron. A variation becomes noticeable in the lower sideband starting around $50^\circ$ when the $d$-wave approaches its node position at $57^\circ$. However, this variation does not lead to a substantial change in the observed angular range. This suggests that none of the cross-channel inter\-ferences outweigh the other interference terms in this case.

Moving to the higher sideband, in accordance with the propensity rule, three-photon absorption in the continuum creates a $g$ electron with notable probability. Therefore, it is reasonable to expect that the angle-dependence of the higher sideband will start to vary early, following the node position of $P_{4}(\theta)$ around $30^\circ$.
Clearly, the significant variation in the RABBITT phase around $50^\circ$ in the higher sideband suggests that cross-channel interferences involving $d$ and/or $g$ electrons start to become comparable and outweigh the remaining interference terms above $50^\circ$. However, our current technique does not allow us to definitively determine which interference term is the most significant.

In the next sideband group, $S_{14}$, we observe that the higher sideband exhibits less angular variation within the observed angular range compared to the $S_{12}$ group.  Additionally, the transition-amplitude ratios to different angular-momentum states change with the energy, potentially shifting the angle where cross-channel interference becomes dominant or, in some cases, nearly eliminating cross-channel interference altogether. Unfortunately, the error bars in the $S_{14,l}$ numbers are too large to draw conclusions with high confidence.

As the probe intensity increases, higher-order transition terms become increasingly significant. In such cases, the transition pathways depicted in Fig.~\ref{fig:Fig5-10a} may no longer provide an adequate description, necessitating the inclusion of higher-order terms.  Furthermore, we note that the angle-dependence of the RABBITT phase changes slowly with varying probe intensity for most sidebands. However, a notable exception is seen in the lower sideband of $S_{12}$, where there is a substantial variation in the angle-dependence of the RABBITT phase with changing probe intensity. This behavior can be attributed to the increasing impact of six-photon transitions involving the under-threshold harmonic ($H_9$) and Rydberg states. These transitions can be intensified by resonances and may also introduce rapid, energy-dependent resonance phases, resulting in a pronounced angle-dependence of the RABBITT phase.

\begin{figure}[t]
\includegraphics[width=0.9\columnwidth]{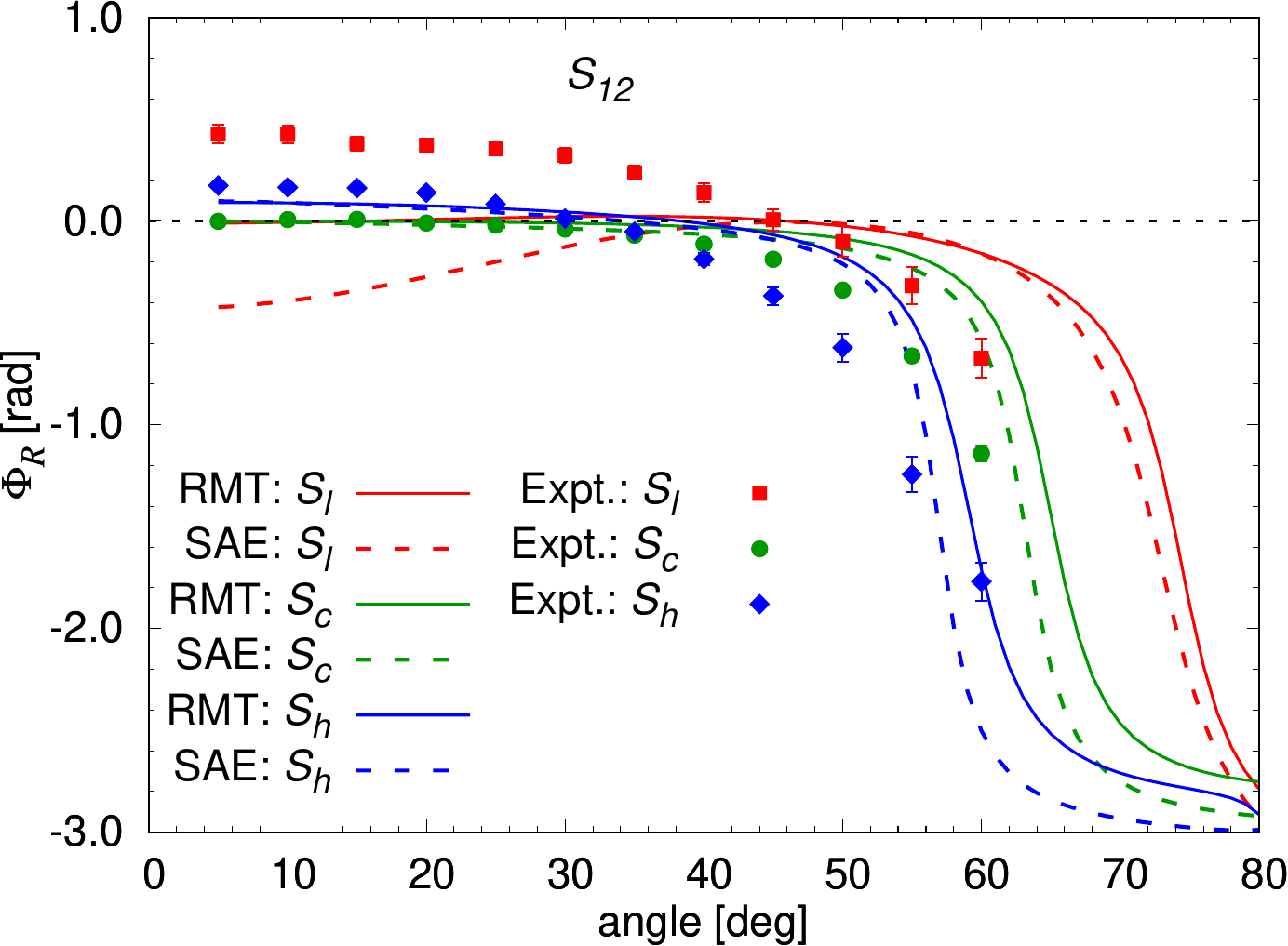}
\includegraphics[width=0.9\columnwidth]{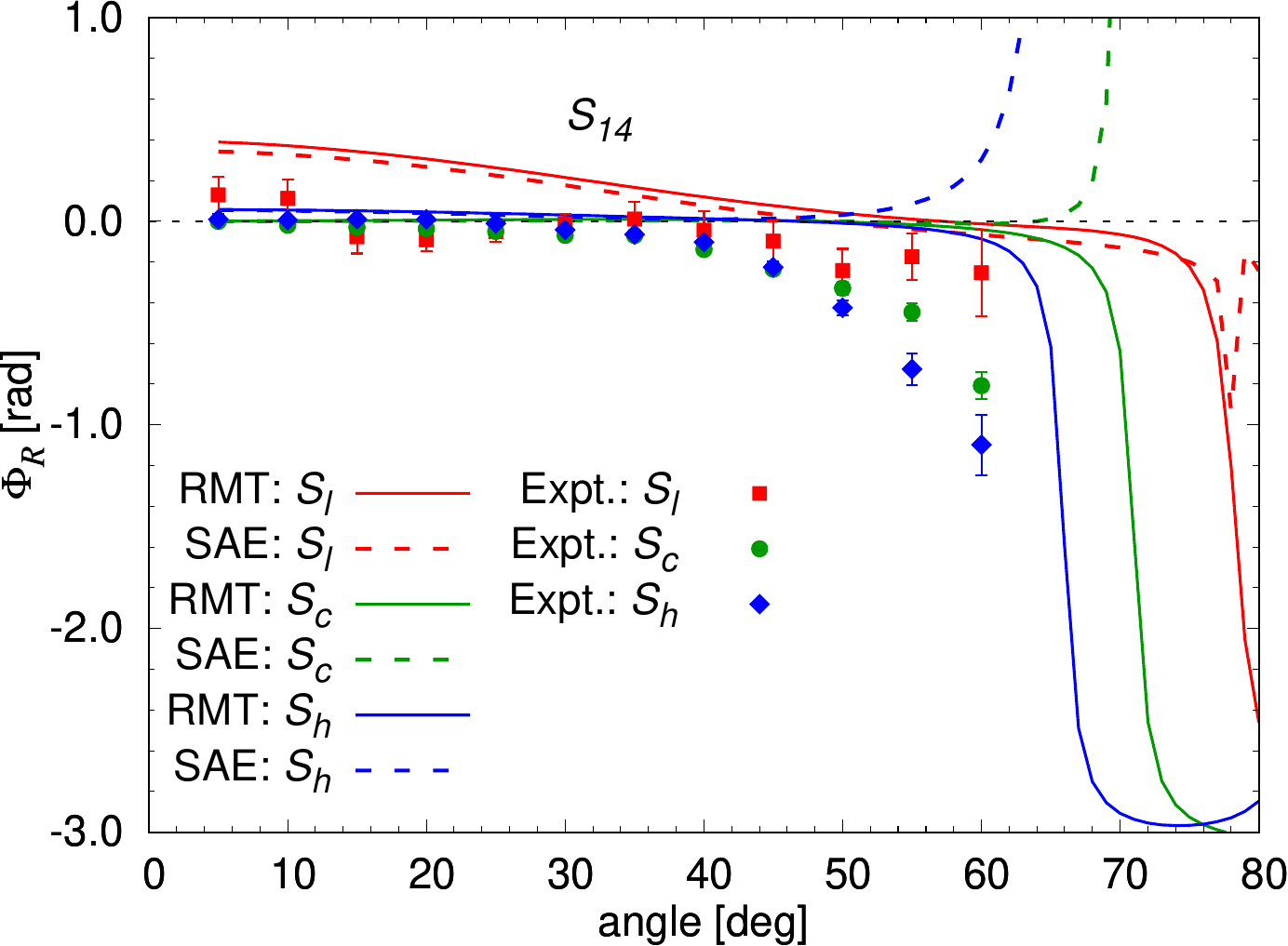}
\caption{SAE and and RMT angle-dependent phase for the $S_{12}$ and $S_{14}$ groups, compared with the experimental data for an IR peak intensity of $I_2 = \rm 7 \times 10^{11}W/cm^2$.} 
\label{fig:Fig-xx}
\end{figure}
 
Figure~\ref{fig:Fig-xx} depicts a comparison between the experimental data and and predictions from the SAE and RMT models for the $S_{12}$ and $S_{14}$ groups.  We only show the case for the peak intensity of $I_2 = \rm 7 \times 10^{11}W/cm^2$, since the findings are similar for the other two.  
Starting with the $S_{12}$ group, we see fairly good agreement between both sets of theoretical predictions and experiment for the center sideband~$S_{12,c}$ and the higher sideband~$S_{12,h}$, whereas there are quantitative differences for $S_{12,l}$.  These include the relative position at small angles: experiment has the phase of $S_{12,l}$ above that of $S_{12,c}$ and $S_{12,h}$, while the SAE model in particular predicts $S_{12,l}$ to start significantly below. 
A possible reason is that higher-order transitions involving Rydberg states are affecting the phase of $S_{12,l}$, and it shifts gradually above that of $S_{12,c}$ as the probe intensity is increased. 
Nevertheless, there is agreement in the drop of the RABBITT phases beyond $\approx 40^\circ$, with $S_{12,h}$ dropping the fastest with increasing angle.

On the other hand, there is substantial disagreement between experiment and theory for the $S_{14}$ group.  While SAE and RMT in general agree well with each other, the experimental data suggest a much stronger angular dependence of all RABBITT phases, while the theoretical predictions are nearly flat. 
A notable exception to the otherwise good agreement between the SAE and RMT predictions is the rapid increase in the SAE RABBITT phase of $S_{14,h}$ at angles beyond $\approx 50^\circ$ and $S_{14,c}$ beyond $\approx 65^\circ$. We note that the signal strength drops fast with increasing angle. Hence, the predictions become very sensitive to the details of the model, in this case the treatment (or the lack thereof) of exchange effects and correlations in the description of the ground state.  While RMT is likely to be superior to SAE in this regard, neither model reproduces the early decreases in the measured phases of $S_{14,c}$ and particularly $S_{14,h}$. Instead, both theories predict the most rapid changes in the angular range where no experimental data are available.  RMT, for example, predicts $S_h$ to drop first around $65^\circ$, followed by $S_c$ around $70^\circ$, and finally $S_l$ around $75^\circ$. 

\bigskip

\section{Conclusions and Outlook} \label{sec:Summary}
In this joint experimental and theoretical study, we extended our previous work on argon and carried out a 3-SB experiment on helium.  This target was chosen for several reasons: 1)~the XUV step is much simpler in helium compared to argon, since only one orbital angular momentum is generated; 2)~we expected a better chance for theory to handle helium rather than the more complex argon target; 3)~there are no known auto\-ionizing resonances in the range of ejected-electron energies studied in the present work.

Indeed, we found that relatively simple numerical models were able to qualitatively, and in some cases also quantitatively, reproduce most of the experimental findings. Exceptions include the angle-integrated phase extracted for the threshold sideband, which is heavily affected by IR-induced transitions involving Rydberg states, and the angle-dependence of the RABBITT phase for one of the two sideband groups studied in this work. 

We hope that the experimental data produced in this study will serve as motivation for future work. This includes a detailed study of the threshold sideband, as well as measurements with improved statistics at lower intensity to reduce higher-order effects.  In addition, we will attempt to increase the cutoff of the higher harmonics in order to be able to compare more sideband groups.
We will also perform a thorough investigation of the sensitivity of theoretical predictions on the details of the model. We are currently extending the RMT calculations to include more coupled states to further improve the bound-state description and even to account for coupling to the ionization continuum in the spirit of the \hbox{$R$-matrix} with pseudo-states (RMPS) approach~\cite{0953-4075-29-1-015}. Since many delays have to be scanned through, such calculations are computationally expensive and have to be planned with great care in light of the available resources.  
\medskip
\begin{acknowledgments}
The experimental part of this work was supported by the DFG-QUTIF program under
Project \hbox{No.~HA 8399/2-1} and \hbox{IMPRS-QD}.
A.T.B., S.S., K.R.H., and K.B.\ acknowledge funding from the NSF through grant
\hbox{No.~PHY-2110023} as well as the  
Frontera Pathways allocation PHY-20028.
A.T.B.\ is grateful for funding through NSERC.
The calculations were performed on Stampede-2 and Frontera at the Texas Advanced Computing Center in Austin (TX).
\end{acknowledgments}
\bibliographystyle{apsrev4-1}

\end{document}